\documentclass[letter]{aa}         % for the letters 
\usepackage{natbib}
\usepackage{graphicx}
\usepackage{txfonts}
\usepackage{hyperref}
\usepackage{amssymb}    % Extra maths symbols
\usepackage{amsmath}    % Advanced maths commands
\usepackage{gensymb}
\usepackage{xcolor}
\usepackage{balance}

\begin{document} 

  \title{Can planetary rings explain the extremely low density of HIP\,41378\,f?}
  %\subtitle{}
  \author{B. Akinsanmi 
          \inst{1,2,6}, N. C. Santos\inst{1,2}, J. P. Faria\inst{1}, M. Oshagh\inst{1,3}, S. C. C. Barros\inst{1}, A. Santerne\inst{4} and S. Charnoz\inst{5}       
          }

  \institute{ Instituto de Astrof\'isica e Ci\^encias do Espa\c co, Universidade do Porto, CAUP, Rua das Estrelas, 4150-762 Porto, Portugal.
\\
  \email{tunde.akinsanmi@astro.up.pt}
 \and
Departamento de F\'isica e Astronomia, Faculdade de Ci\^encias, Universidade do Porto, Rua do Campo Alegre, 4169-007 Porto, Portugal.
 \and  Institut f\"ur Astrophysik, Georg-August-Universit\"at G\"ottingen, Friedrich-Hund-Platz 1, 37077 G\"ottingen, Germany
 \and Aix Marseille Univ, CNRS, CNES, LAM, Marseille, France.
 \and Institut de Physique du Globe de Paris (IPGP), 1 rue Jussieu, 75005, Paris, France
 \and National Space Research and Development Agency. Airport Road, Abuja, Nigeria.
 }

  \date{Received ; accepted }

  \abstract{The presence of rings around a transiting planet can cause its radius to be overestimated and lead to an underestimation of its density if the mass is known. We employed a Bayesian framework to show that the anomalously low density ($\sim$0.09\,g\,cm${^{-3}}$) of the transiting long-period planet HIP\,41378\,$f$ might be due to the presence of opaque circum-planetary rings. Given our adopted model priors and data from the K2 mission, we find the statistical evidence for the ringed planet scenario to be comparable to that of the planet-only scenario. The ringed planet solution suggests a larger planetary density of $\sim$1.23\,g\,cm$^{-3}$ similar to Uranus. The associated ring extends from 1.05 to 2.59 times the planetary radius and is inclined away from the sky plane by $\sim$25\degree. Future high-precision transit observations of HIP\,41378\,$f$ would be necessary to confirm/dismiss the presence of planetary rings.}

\titlerunning{Rings around HIP\,41378 f}
\authorrunning{B. Akinsanmi et al.}

\maketitle

%________________________________________________________________

\section{Introduction}

Planetary rings are exciting features yet to be detected around exoplanets despite their prevalence around the giant planets and other rocky bodies of the solar system. A number of studies have proposed methods to identify and characterise their signatures from transit light curves, Rossiter-McLaughlin (RM)\ signals, and reflected light signals (e.g. \citealt{barnes, ohta, mooij, santos}).\\

The transit method is very attractive for probing the presence of rings as they cause a number of effects in the transit light curve \citep{barnes,tsunski}. Searches for rings in transit data have thus been performed and in some cases possible ring signals have been identified or constraints placed on ring parameters (e.g. \citealt{ken, heising,aizawa,aizawa18}). The presence of rings around a transiting planet would cause a deeper transit signal which could be mistaken to be due to a larger planetary radius \citep{akin18}. The overestimated radius leads to an underestimation of the density of a planet if its mass is known \citep{zuluaga}.\\

Extremely low-density planets, so-called super-puffs, thus provide a unique and unexplored planet class to search for the presence of rings \citep{piro19}. Prime examples of these super-puff planets are  Kepler-51\,$b$, $c,$ and $d$ \citep{masuda14} and  Kepler-79\,$d$ \citep{jontoff}, which all have densities below 0.1\,g\,cm$^{-3}$. However, the low signal-to-noise data due to their faint stars makes them unsuitable for probing the transit signature of rings.

Interestingly, the bright star HIP\,41378 (K=7.7\,mag), which was observed in campaigns C5 and C18 of the K2 mission has been shown to host at least five transiting planets \citep{Vanderburg_2016}. In particular, HIP\,41378\,$f$ was found to have a period of 542\,days and a mass of $12\pm3\,M_{\oplus}$ \citep{santerne}. Combining this mass with the derived planetary radius of $9.2\pm0.1\,R_{\oplus}$ gives an anomalously low planetary density of $\sim$0.09\,g\,cm${^{-3}}$ (Table \ref{parameters}), which puts it in the class of super-puffs.

We therefore investigate the possibility that the low density of HIP\,41378\,$f$ can be due to the presence of planetary rings. Long-period planets, such as HIP\,41378\,$f$ with semi-major axis of $\sim$1.4\,AU, are particularly interesting in the search for rings as they can be similar to the ringed objects in the solar system which all orbit far from the Sun. At large distances from their host stars, planets are less influenced by the tidal forces of the star. This allows the planets to have large enough Hill radii to support rings and the rings are able to have a wide variety of orientations that can favour their detection \citep{sch11}. The orbit of HIP\,41378\,$f$ is consistent with an eccentricity, $e$, of zero \citep{santerne}, which is also favourable for hosting stable rings as it ensures a  constant stellar tidal influence.

In this Letter, we perform Bayesian model comparison between a ringed planet scenario and the planet-only scenario to determine which of these scenarios is most probable given the data.

\section{Transit data and model priors}
\label{sect2}
\subsection{Models}
\begin{figure}[tp!]
\centering
\includegraphics[width=1\linewidth]{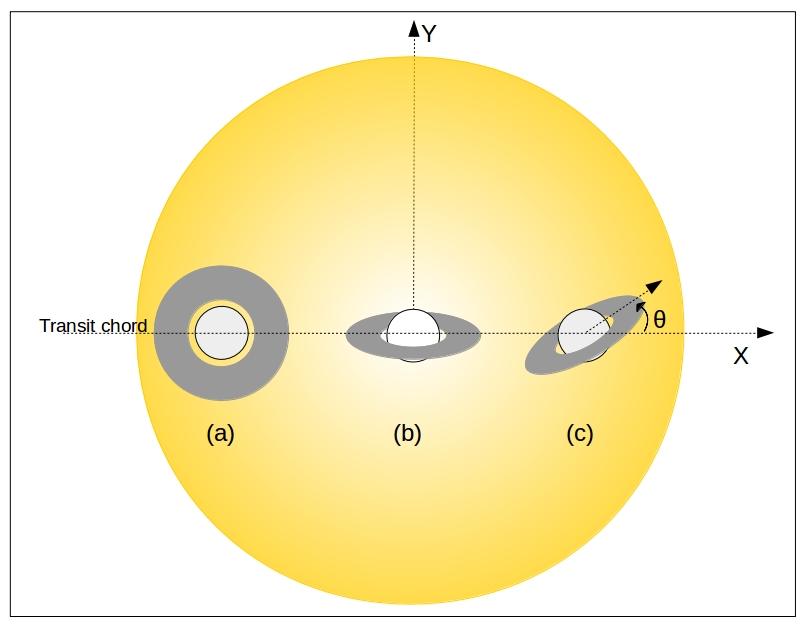}
\caption{ Schematic of ringed planet transit with multiple ring orientations with sky plane XY. (a) Planet with face-on ring ($i_r=0\degree$); (b) planet with $i_r=60\degree$, $\theta=0\degree$; and (c) planet with $i_r=60\degree$, $\theta=30\degree$.}
\label{schematic}
\end{figure}

We model the photometric transit of a ringed planet using \texttt{SOAP3.0} \citep{akin18}. The ring is defined by an inner and outer radii $R_{in}$ and $R_{out}$ in units of the planetary radius $R_{p}$ with constant opacity $\tau$. The ring has two orientation angles: $i_{r}$ is the inclination of the ring plane with respect to the sky plane ($0\degree$ and $90\degree$ for face-on and edge-on rings projections, respectively), while $\theta$ defines the obliquity/tilt of the ring from the orbital plane (measured anti-clockwise from the transit chord; see Fig. \ref{schematic} and also \citealt{akin18}). The planet-only model has the usual spherical model transit parameters. A description of the relevant parameters for both models is given in Table \ref{priors}. To investigate the ringed planet hypothesis, we perform a Bayesian model comparison by computing the evidence (see Sect. \ref{sect3} ) for the planet-only and ringed planet scenarios given the observational data from the K2 mission. 
\subsection{Transit data}

\begin{figure}[tp!]
\centering
\includegraphics[width=1\linewidth]{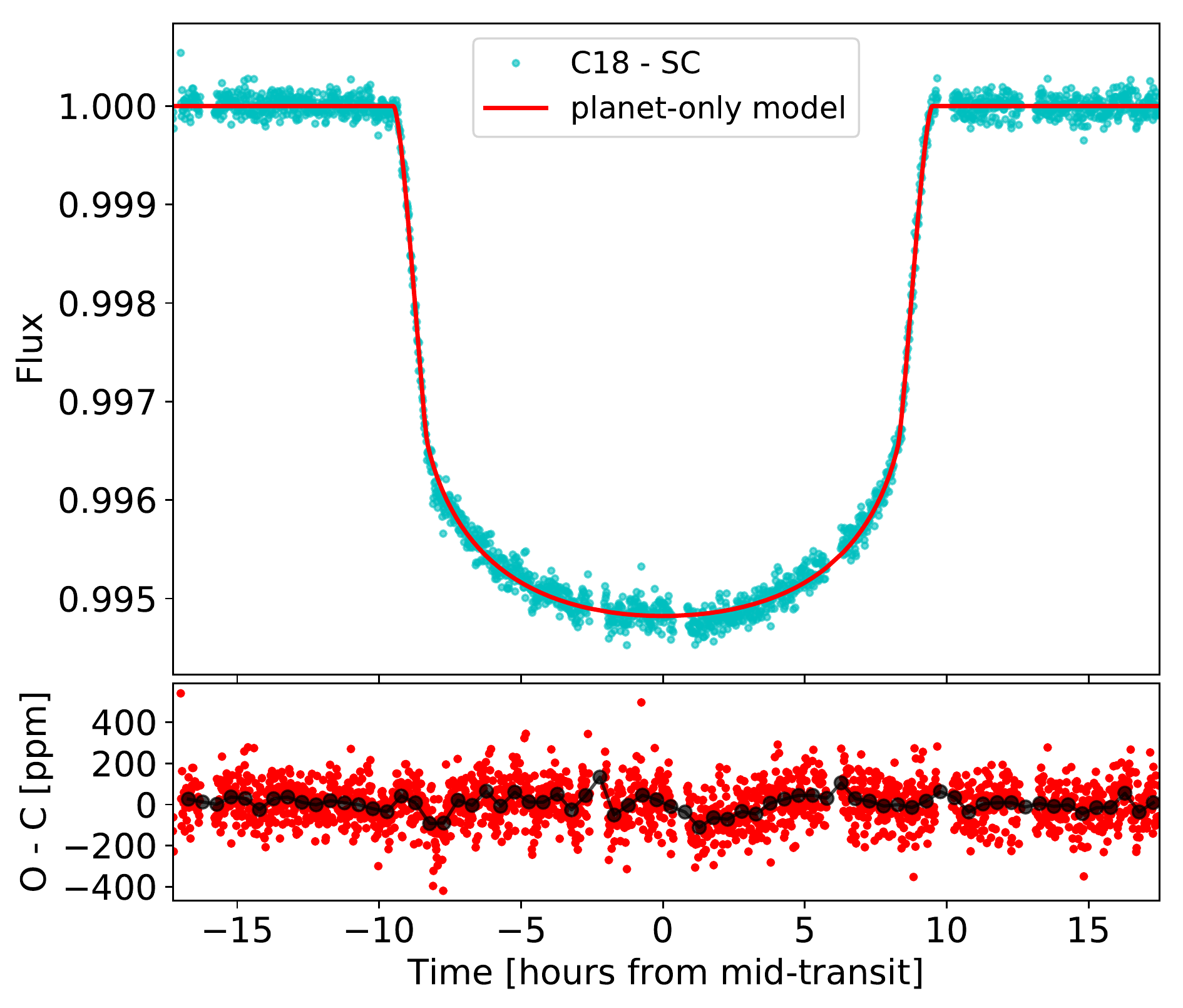}
\caption{ Spherical planet transit model fit (red line) to the C18 short-cadence data (cyan points) of  HIP\,41378\,$f$ and the residual (red points). The 30\,min binned residuals (black) is overplotted on the residuals.}
\label{lc}
\end{figure}

The star HIP\,41378 was observed in long-cadence mode (LC) during K2 C5 and then in short-cadence mode (SC) in C18. We used the reduced HIP\,41378 light curves from \citet{santerne}, which were produced with the K2SFF pipeline \citep{Vanderburg14} without significant modification of the in-transit data. Searching for ring signatures in light curves requires high time resolution data, so we performed our analyses on the C18 SC light curve of HIP\,41378\,$f$ (1933 transit data points) and checked the consistency of the result with the C5 light curve. A cursory fit of a spherical planet transit model to the light curve (Fig. \ref{lc}) reveals no visual sign of the characteristic residual ingress and egress anomalies that can be caused by the presence of rings\footnote{Although we noticed some artefacts of the reduction process in the C18 light curve of HIP\,41378\,$f$, we chose not to perform further corrections to prevent the removal of possible ring features.} \citep{akin18}. However, it has been shown that these ring signals can be masked if $R_{in}$ is sufficiently close to the planet surface \citep{ohta}. The lack of discernible ingress and egress signature in the residual could also imply that any possible ring around the planet that is capable of producing the observed transit depth must be densely packed and opaque or else the transition between the less opaque ring and completely opaque planet would have left a significant imprint during ingress and egress. Therefore, we assume that the putative ring is completely opaque. 

\subsection{Model priors}

To calculate the evidence of each model given the C18 SC data, it is important to define appropriate priors on the parameters of the models as the evidence is very sensitive to their values. 
The prior on the scaled semi-major axis, $a/R_{\ast}$, is obtained using Kepler's third law with values of the planetary period and the stellar density (Table \ref{parameters}). A careful selection of priors for the stellar limb darkening coefficients (LDCs) is necessary since their effect is prominent at ingress and egress where ring signatures can also manifest themselves. The quadratic LDCs ($u_1$, $u_2$) were first interpolated from \citet{claret} using parameters of the host star \citep{lund}. Thereafter, a better estimate of their values was obtained from the joint transit fitting of the other planets in this system (excluding planet $f$). The resulting values and associated uncertainties were then used as priors in both the planet-only and ringed planet models (see Table \ref{priors}). The planet eccentricity was kept fixed at zero as derived in \citet{santerne}.

To define priors for the planetary radius $R_{p}$, we consider the radius distribution of detected planets\footnote{\href{https://exoplanetarchive.ipac.caltech.edu/}{https://exoplanetarchive.ipac.caltech.edu/}} with masses within $3\sigma$ of the mass of HIP\,41378\,$f$. This broad distribution is used because it spans a wide range of planetary radii including those of the aforementioned super-puff planets making it suitable as prior for the planet-only and ringed planet models. Given the mass, HIP\,41378\,$f$ is expected to be a gaseous planet so we remove planets with radii below $2\,R_{\oplus}$ to avoid planets that are consistent with rocky compositions \citep{Marcy_2014}. The resulting radius distribution was found to be well represented by a log-normal distribution (see Fig. \ref{rad_distr}), which was then used as the prior on $R_{p}$ in both models.

To obtain priors for the outer ring radius, $R_{out}$, we consider that rings are only stable within the Roche radius of the planet. Beyond this radius, the ring materials are unstable and ultimately coalesce to form satellites. The Roche radius is given by \begin{equation}
\label{roche}
    R_{Roche} = 2.45\,R_{p} \left( \frac{\rho_{p}}{\rho_{r}} \right)^{1/3}.
\end{equation}
Therefore, the possible rings around this planet must have $R_{out}$\,$\leq$\,$R_{Roche}$. However, the underlying planet density $\rho_p$ and ring density $\rho_r$ required to calculate $R_{Roche}$ are unknown. The main rings of the giant planets of the solar system are within the Roche radius of their respective planet, which does not vary much between planets and is found to be generally around $2-3\,R_p$ \citep{Charnoz2018}. We adopt the upper limit and assume that the possible rings around this planet are also within $R_{Roche}=3\,R_p$. We assume that the rings can possibly extend from the planet surface so we adopt uniform priors on $R_{out}$ from 1\,$R_{p}$ to 3\,$R_{p}$. Since we must have $R_{in}\leq R_{out}$, the priors on $R_{in}$ is from $1\,R_{p}$ to $R_{out}$ ; the value of $R_{out}$ is updated at every iteration of the computation. For a planet to host rings with bound stable orbits, its Roche radius has to be within two-thirds of its Hill radius $R_{H}$ \citep{sch11}. We derive $R_{H} = 180\,R_{p}$ for HIP\,41378\,$f$ (i.e.\,$R_{H} \gg R_{Roche}$), implying that it can host stable and long-lived rings. 

Given that the equilibrium temperature of this planet, $T_{eq}$\,$\simeq$\,294\,K, is higher than the melting temperature of water ice, the materials of any ring around this planet needs to have higher melting temperatures and densities than ice ($\rho_r > 1$\,g\,cm$^{-3}$; $2-5$\,g\,cm$^{-3}$ for rocky rings). Therefore, our computation enforced that the proposed solution must have $\rho_r$\,>1\,g\,cm$^{-3}$.

The ring inclination $i_r$ ranges from 0\degree\,(face-on) to 90\degree (edge-on). We note that at edge-on the ring has no effect on the light curve as its projected area is negligible. The projected area of the ring is proportional to the cosine of $i_{r}$, so we use a prior distribution which is uniform in $\cos\,i_{r}$. We use an uninformative uniform prior on the ring obliquity, $\theta$, ranging from 0 - 180\degree.

We note that different assumptions from those stated above regarding the parameters of the models could change the resulting evidence for the models and also lead to a different ring solution. Nevertheless, we adopted these priors as they are physically representative of the current knowledge of planets and rings.

%______________________________________________________________

\section{Model comparison}
\label{sect3}
We apply a Bayesian framework (see Appendix \ref{evidence_calc}) to compare the log evidence for the ringed planet model ($\log \mathcal{Z}_{R}$) to that for the planet-only model  ($\log \mathcal{Z}_{pl}$) using the Bayes factor given by \begin{equation}
    K= \exp{(\log \mathcal{Z}_{R}\,-\,\log \mathcal{Z}_{pl})}.
    \label{bayes_factor}
 \end{equation}
For $1<K<3.2$, the ringed model is barely more probable than the planet-only model, whereas $K>3.2$ implies substantial evidence against the planet only model \citep{kass}. 

We compute the evidence for the planet-only model with $R_{p}$, $a/R_{\ast}$, $i_{p}$, $u_{1}$, and $u_{2}$ as free parameters while the ringed planet model additionally has $R_{in}$, $R_{out}$, $i_{r}$, and $\theta$. These parameters and their adopted priors are described in Table \ref{priors}. The same priors are used when both models have parameters in common. The results are given in Table \ref{result_table} and the posteriors of the parameters from both models are shown in Fig. \ref{posteriors}.

\begin{table}[t!]
\centering
\caption{Performance of the models: Bayesian evidences $\log \mathcal{Z}$ and maximum log-likelihoods $\log \hat{\mathcal{L}}_{\Theta}$. The median of posterior samples for each model is also given alongside the 68\% credible interval.}
\label{result_table}
\begin{tabular}{lll}
\hline \hline
Parameter   \qquad \qquad     & Planet-only model         & Ringed model \\
\hline
$\log \mathcal{Z}$               & $14952.44$    \qquad \qquad \qquad           & $14952.85$    \\[1pt]
$\log \hat{\mathcal{L}}_{\Theta}$ & $14970.85$                       & $  14972.60$           \\[1pt]
%BIC                             & -29939.986                         &   -29918.396       \\[3pt]
$R_{P}\,[R_{\oplus}]$             & $9.21\pm0.01$                  &   $3.7^{+0.3}_{-0.2}$                  \\[3pt]
$a/R_{\ast}$                    & $231.6\pm0.7$                    &  $231.0\pm0.6$               \\[4pt]
$i_{p}\,[\degree]$              & $89.97\pm0.01$                     &   $89.97\pm0.01$                 \\[4pt]
$u_{1}$                                                 & $0.32\pm0.01$                                 & $0.32\pm0.01$   \\[4pt]
$u_{2}$                                                 & $0.28\pm0.01$                                 & $0.28\pm0.01$   \\[4pt]
$R_{in}\,[R_{p}]$               & -                                  &  $1.05^{+0.05}_{-0.03}$                    \\[4pt]
$R_{out}\,[R_{p}]$              & -                                  &   $2.6\pm0.2$                 \\[4pt]
$i_{r}\,[\degree]$              & -                                  &  $25^{+3}_{-4}$                   \\[4pt]
$\theta\,[\degree]$             & -                                  & $95^{+16}_{-17}$                   \\[4pt]
$\rho_p\,[g\,cm^{-3}$]          & $0.09\pm0.02$                                         & $1.2\pm0.4$  \\[4pt]
\hline 
\end{tabular}%
\end{table}

Comparing the evidence for both models using eq. \ref{bayes_factor} results in a Bayes factor $K=1.51$ in marginal favour of the ringed planet model \citep{kass}. Because the value of K is close to 1, this implies that given the K2 C18 SC data and the adopted model priors, the ringed planet scenario is not significantly more probable and only provides a comparable evidence to the planet-only scenario. This is not surprising given that the characteristic ingress and egress transit signatures of rings are either absent or well suppressed in the data making the light curves of both models similar. It is however interesting that the ringed model has comparable evidence to the planet-only model despite the introduction of four extra parameters, which increases the prior volume compared to the planet-only model. 

As previously mentioned, model comparison using Bayes factor is sensitive to the adopted priors for the models hence our selection of priors that are as physical as possible. For example, the adopted prior radius distribution favours smaller planet sizes but this is indeed the case given the measured mass of the planet. Not taking into account the knowledge of radius distribution would lead to a result that favours the planet-only model. Also, deriving the adopted radius distribution from planets with masses within $1\sigma$ of the mass of HIP\,41378\,$f$ instead of $3\sigma$ leads to a prior on $R_{p}$ that only favours the ringed planet model. 

The resulting ringed planet solution suggests a smaller planetary radius of $R_p=3.7^{+0.3}_{-0.2}\,R_{\oplus}$, which is in the radius range obtained using mass-radius prediction tools such as \texttt{forecaster}\footnote{\href{https://github.com/chenjj2/forecaster}{github.com/chenjj2/forecaster} \citep{chen}} ($3.3\pm1.4\,R_{\oplus}$) and \texttt{bem}\footnote{\href{https://github.com/soleneulmer/bem}{github.com/soleneulmer/bem} \citep{moll}} ($3.8\pm0.4\,R_{\oplus}$). Combining this radius with the planet mass gives a higher planetary density of $\rho_{p}$\,=\,1.2$\pm0.4$\,g\,cm$^{-3}$ similar to that of Uranus ($1.27$\,g\,cm$^{-3}$). The associated ring begins close to the planet surface with $R_{in}$\,=\,$1.05\,R_{p}$ and extends to $R_{out}$\,=\,$2.59\,R_{p}$. Although Saturn's fairly transparent D ring also begins close to the planet at $1.11\,R_{p}$, it is unclear if dense opaque rings can have such proximity to the planet. We calculate the density of the possible ring materials that can be sustained within the obtained $R_{out}$ by setting $R_{out}=R_{Roche}$ in eq. \ref{roche}. We obtain $\rho_r=1.08\pm0.3$\,g\,cm$^{-3}$ with 95\% upper limit of 1.63\,g\,cm$^{-3}$, which is denser than water ice but not as dense as typical rocky ring materials. The plausibility of such low-density ring particles is questionable at the equilibrium temperature of the planet. Although porous rocky materials can have such low densities (below 2\,g\,cm$^{-3}$) as measured for some asteroids \citep{carry}, the possible formation scenario  for such a ring is unknown.

\begin{figure}[t!]
    \includegraphics[width=1.1\linewidth]{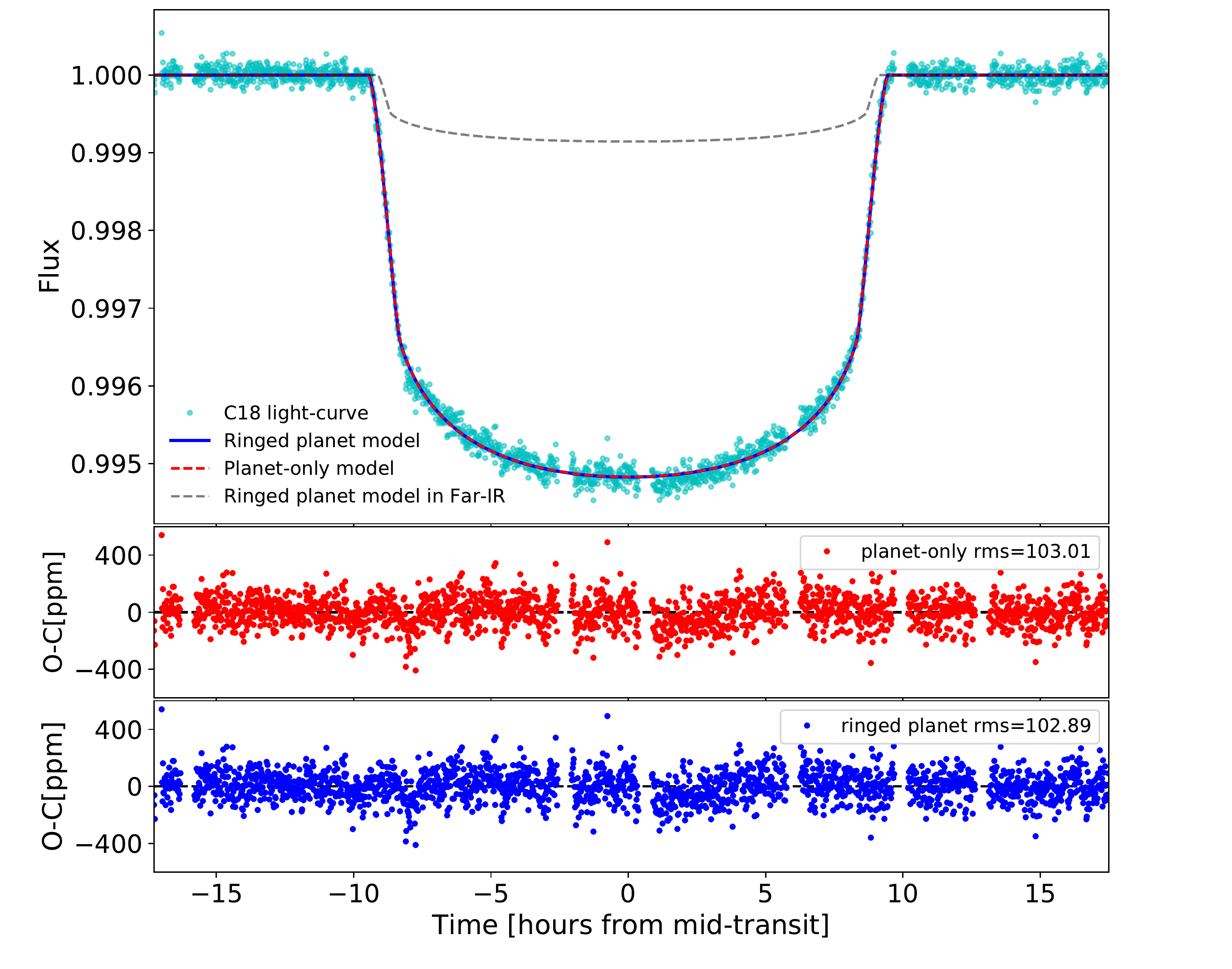}
    \caption{Fit of the planet-only (red dashed line) and the ringed planet (blue solid line) to the C18 SC data (cyan points). The residuals of same colour are also shown with the  rms value in ppm. Also plotted in the top panel (grey dashed line) is the predicted light curve when the ringed planet is observed in the FIR where rings are expected to be transparent.}
    \label{fit}
\end{figure}

Given the adopted model priors, the best ringed planet solution gives a ring inclination $i_{r}$\,=\,25\degree\,, which allows sufficient ring projected area to match the observed transit depth. The 95\% upper limit on $i_r$ is 30\degree. So for randomly orientated ring inclinations, the statistical probability of finding a ring with $i_{r}$ lower than $30\degree$ is $\mathcal{P}=1-\cos\,(30\degree)\simeq 13\%$, which is high considering that the probability of transit for this planet is only $\sim$0.5\%. 

We can determine the plane in which the putative ring lies (see Appendix \ref{ringplane}) by computing the ratio of the Laplace radius to the Roche radius, $R_{L}/R_{Roche}$,  given in eq. \ref{laplace} . Assuming quadrupole moment values, $J_2$, in the range of the solar system giant planets (0.003 - 0.1), we obtain $R_{L}/R_{Roche}>1.7$, implying that the plane of the possible ring around this planet aligns with the equatorial plane of the planet \citep{sch11}. Since the ring solution indicates a ring tilted by $\theta\simeq95\degree$ from the orbital plane, it implies that the equatorial plane of the planet is also $95\degree$ from the orbital plane similar to Uranus (97.86\degree).

\begin{figure}[t!]
    \includegraphics[width=1.1\linewidth]{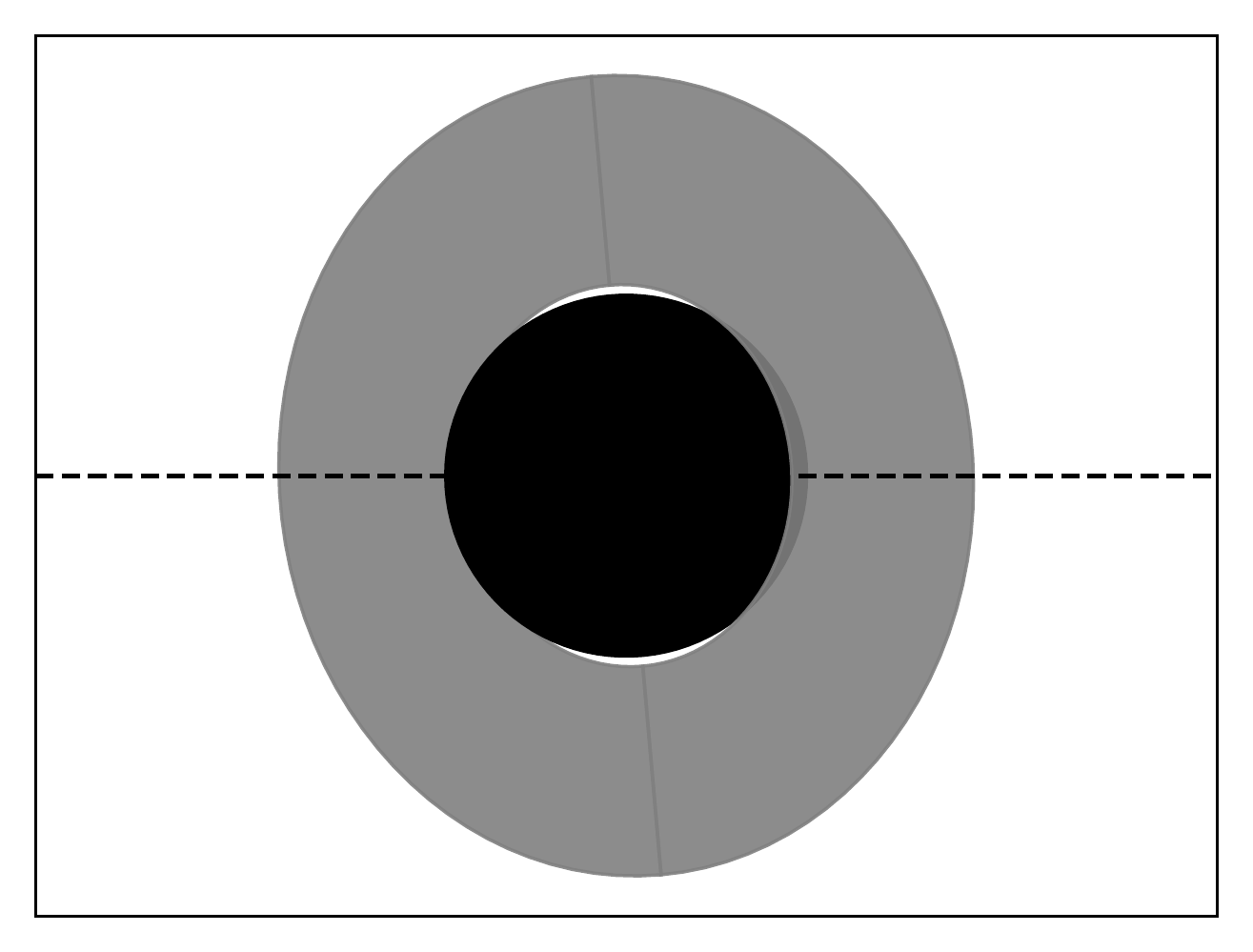}
    \caption{Schematic of the ringed planet solution with $i_r=25\degree$ and $\theta$\,=\,$95\degree$. The dashed line indicates the transit chord.}
    \label{ring_solution}
\end{figure}

The fit to the data using the best parameters from  both models is shown in Fig. \ref{fit}. It is seen from the root mean square (rms) of the residuals that  both models provide comparable fit to the data. This indicates that the possible ring around this planet emulates well the signal of a planet-only model, thereby making it difficult to distinguish between both models. As a consistency check, we performed a fit of both models to the K2 C5 LC light curve (see Fig. \ref{c5}) and found that the resulting values of the parameters agree with our results from the C18 light curve within $1\sigma$. A schematic of the ringed planet solution is shown in Fig. \ref{ring_solution}.

\section{Discussion and conclusions}
\label{sect4}

A smaller planet with opaque rings provides not only a good fit to the K2 light curve of HIP\,41378\,$f$ but can also explain its unusually low density. Nevertheless, it is possible that other phenomena may also be able to explain the anomalous radius/density. For instance, it is possible that the observed large radius is due to the planet having a small core and an extended atmosphere, likely composed of hydrogen. Super-Earths with masses up to $10\,M_{\oplus}$ are capable of having such hydrogen-rich atmospheres that may dramatically increase the planet radius \citep{miller}. \citet{adams} found that an atmosphere with 10\% the mass of a planet can cause its radius to increase by up to 60\%. This is especially the case if the atmosphere is undergoing hydrodynamic loss (outflows) owing to the low surface gravity of the planet \citep{wang}. These outflows carry dust to high altitudes (enhancing the opacity of the atmosphere), which inflates the observed radius of the planet and even leads to featureless transmission spectra when probing the atmospheres. However, these outflows seem to affect planets with masses lower than $10\,M_{\oplus}$, which have weak gravitational wells and so it is not clear if they can occur in higher mass planets such as HIP\,41378\,$f$.

Several studies have also provided some explanations for the radius inflation of exoplanets mostly pointing to the correlation between the radius inflation and the level of radiation it receives from the star \citep{lopezfortney}. For a particular star, the planets in close proximity generally receive higher stellar insolation and are more inflated than those further out. At the distance of 1.4\,AU, HIP\,41378\,$f$ receives only a low level of irradiation that is not sufficient to significantly puff it up as observed. Although young planets ($<$10\,Myrs) are also expected to be inflated because of retained internal heat from their formation, this might not explain the case of HIP\,41378\,$f$ as it is estimated to be 3.1Gyrs old \citep{lund} and is expected to have cooled off.

Besides focussing on the enlarged radius, it is necessary to check the possibility that the derived mass for the planet is not underestimated. The induced Radial Velocity (RV) signal amplitude ($\sim$1\,m\,s$^{-1}$) of the planet is at the level of the instrumental stability and thus the derived mass could be influenced by unknown systematics \citep{santerne}. However, a larger planetary mass is unlikely as it would cause larger RV amplitudes which would have been easier to detect. Further RV observations of this target using high-precision spectrographs have been suggested in order to refine the planetary mass \citep{santerne}.

Having considered these non-exhaustive alternatives, we conclude that the ring hypothesis presents a possible option to explain the observed low density. Further observations will be necessary to confirm/characterise the ring scenario. Transmission spectroscopy can be useful in probing the nature/presence of such rings as their opacity might vary with wavelength depending on the composition and density of the ring materials. However, solar occultations of Saturn's main rings have revealed featureless transmission spectra in which the ring materials are almost completely opaque at visual and near-infrared wavelengths \citep{nicholson08}. At far-infrared (FIR) wavelengths, the rings should be optically thinner and we might expect to measure a shallower transit corresponding to a smaller planetary radius. The predicted light curve of the ringed planet at the FIR wavelength (where the ring might be transparent) is also shown in Fig \ref{fit}. Additionally, RM measurements \citep{gaudi_winn} during the transit can be used to probe the presence of rings around the planet (see Appendix \ref{appendix_rm}).

As the Bayesian evidence for the ringed planet model is comparable to that of the planet-only model, it is difficult to categorically ascertain the reality of these rings as they mimic well the light curve of a planet-only model. Thus, we are only able to say, given the data, that the ring hypothesis presents one plausible explanation for the inferred low density of the planet. The ringed planet scenario also poses a challenge regarding the possibility of hosting low-density/porous ring materials at the high equilibrium temperature of the planet. This planet will benefit from future transit observations to validate its true nature. Transit observations with higher precision (e.g. using the Hubble Space Telescope or James Webb Space Telescope) will be necessary to identify ingress and egress signatures which will be useful in constraining the parameters of the possible ring and the underlying planet radius.

\begin{acknowledgements}
This work was supported by Funda\c c\~ao para a Ci\^encia e a Tecnologia (FCT, Portugal)/MCTES through national funds by FEDER through COMPETE2020-POCI by these grants: UID/FIS/04434/2019, PTDC/FIS-AST/28953/2017\,\&\,POCI-01-0145-FEDER-028953, and PTDC/FIS-AST/32113/2017 \& POCI-01-0145-FEDER-032113. BA acknowledges support from the FCT PhD programme PD/BD/135226/2017. S.C.C.B. acknowledges FCT support through IFCT contracts IF/01312/2014/CP1215/CT0004. MO acknowledges support of the DFG priority program SPP1992 "Exploring the Diversity of Extrasolar Planets (RE1664/17-1)". MO, BA, JF, also acknowledge support from the FCT/DAAD bilateral grant 2019 (DAAD ID:57453096).

\end{acknowledgements}

\bibliographystyle{aa} % style aa.bst
\bibliography{references} % your references Yourfile.bib

\begin{appendix}
 \section{}
\begin{table}[h!]

        \caption{Parameters of HIP\,41378 star and planet $f$ \citep{lund, santerne}.}
        \label{parameters}
        \centering
        
        \begin{tabular}{lll}
                \hline \hline
                Parameter [unit]              &Symbol                         & Value              \\
                \hline
                Stellar mass $[M_{\odot}]$          &   $M_{\ast}$                      & $1.160\pm0.04$     \\[3pt]
                Stellar radius $[R_{\odot}]$        &   $R_{\ast}$                      & $1.273\pm0.02$    \\[3pt]
                Stellar density $[\rho_{\odot}]$   &   $\rho_{\ast}$                   & $0.563\pm0.01$    \\[3pt]
                Effective temperature [K]  \qquad   &   $T_{\mathrm{eff}}$              & $6320^{+60}_{-30}$ \\[3pt]
                Stellar rotation velocity $[km\,s^{-1}]$                & $\nu\sin i_{\ast}$                              & $5.6\pm0.5$ \\[3pt]
                Planet period [days]                &   $P$             \qquad          & $542.08$
                \\[3pt]
                Transit time [BJD]                  & $t_{0}$                           & $2457186.91$
                \\[3pt]
                Planet mass $[M_{\oplus}]$          &   $M_{p}$                         & $12\pm3$
                \\[3pt]
                Planet radius $[R_{\oplus}]$        &   $R_{p}$                         & $9.2\pm0.1$
                \\[3pt]
                Planet density $[g\,cm^{-3}]$           &   $\rho_{p}$                      & $0.09\pm0.02$
                \\[3pt] 
                Inclination [\degree]               &   $i_{p}$                         & $89.97\pm{0.01}$
                \\[3pt]
                Semi-major axis                     &   $a/R_{\ast}$                    & $231.1\pm 0.8$
                \\[3pt]
                Equilibrium temperature [K]          & $T_{\mathrm{eq}}$    \qquad            & $294^{+3}_{-1}$ 
                \\[1.5pt]
                \hline
        \end{tabular}%
        
\end{table}

\begin{table}[h!]
        \centering
        \caption{Description and adopted priors on the parameters of the planet-only and ringed planet models. $^+$\,specifies parameters with the same priors in both models. The notation $\mathcal{N}(a,b)$ refers to a normal prior with mean $a$ and standard deviation $b$, $\mathcal{U}(a,b)$ refers to a uniform prior between $a$ and $b$, $\mathcal{F}(a)$ refers to a parameter fixed to value $a$, while $\log\mathcal{N}(s,a,b)$ refers to a log-normal prior with shape parameter $s$ shifted and scaled by $a$ and $b,$ respectively. }
        \label{priors}
        \begin{tabular}{lll}
                \hline \hline
                Parameter    \qquad                      & Description     \qquad \qquad  & Prior \\
                \hline
                $R_{p}\,[R_{\ast}]\,^{+}$                       & Planet radius              & $\log \mathcal{N}$(0.95,1.88,1.09)                 \\[3pt]
                $a/R_{\ast}\,^{+}$                              & Semi-major axis             & $\mathcal{N}$(231.07, 0.76)                           \\[3pt]
                $i_{p}\,[\degree]\,^{+}$                & Inclination of orbit        & $\mathcal{U}(\cos 90, \cos 89.9 )$       \\[3pt]
                $e\,^{+}$                                                               & Eccentricity                              & $\mathcal{F}(0)$  \\[3pt]
                $u_{1}$, $u_{2} \,^{+}$                                 & \begin{tabular}[c]{@{}l@{}}Limb darkening\\ coefficients\end{tabular}
                      & \begin{tabular}[c]{@{}l@{}}$\mathcal{N}$(0.307, 0.006),\\  $\mathcal{N}$(0.31, 0.02) \end{tabular}
 \\[8pt]
                $R_{out}\,[R_{p}]$                      & Outer ring radius           & $\mathcal{U}$(1.0, 3.0)                  \\[3pt]            
                $R_{in}\,[R_{p}]$           & Inner ring radius                       & $\mathcal{U}$(1.0, $R_{out}$)             \\[3pt]
                $i_{r}\,[\degree]$          & Ring inclination                       & $\mathcal{U}(\cos 90, \cos 0)$           \\[3pt]
                $\theta\,[\degree]$         & Ring obliquity                          & $\mathcal{U}$($0$, $180$)          \\[3pt]     
                \hline
        \end{tabular}%  
%            \vspace{1ex}
 %    {\raggedright \\ \par}
\end{table}

\begin{figure}[h!]
\centering
\includegraphics[width=0.9\linewidth]{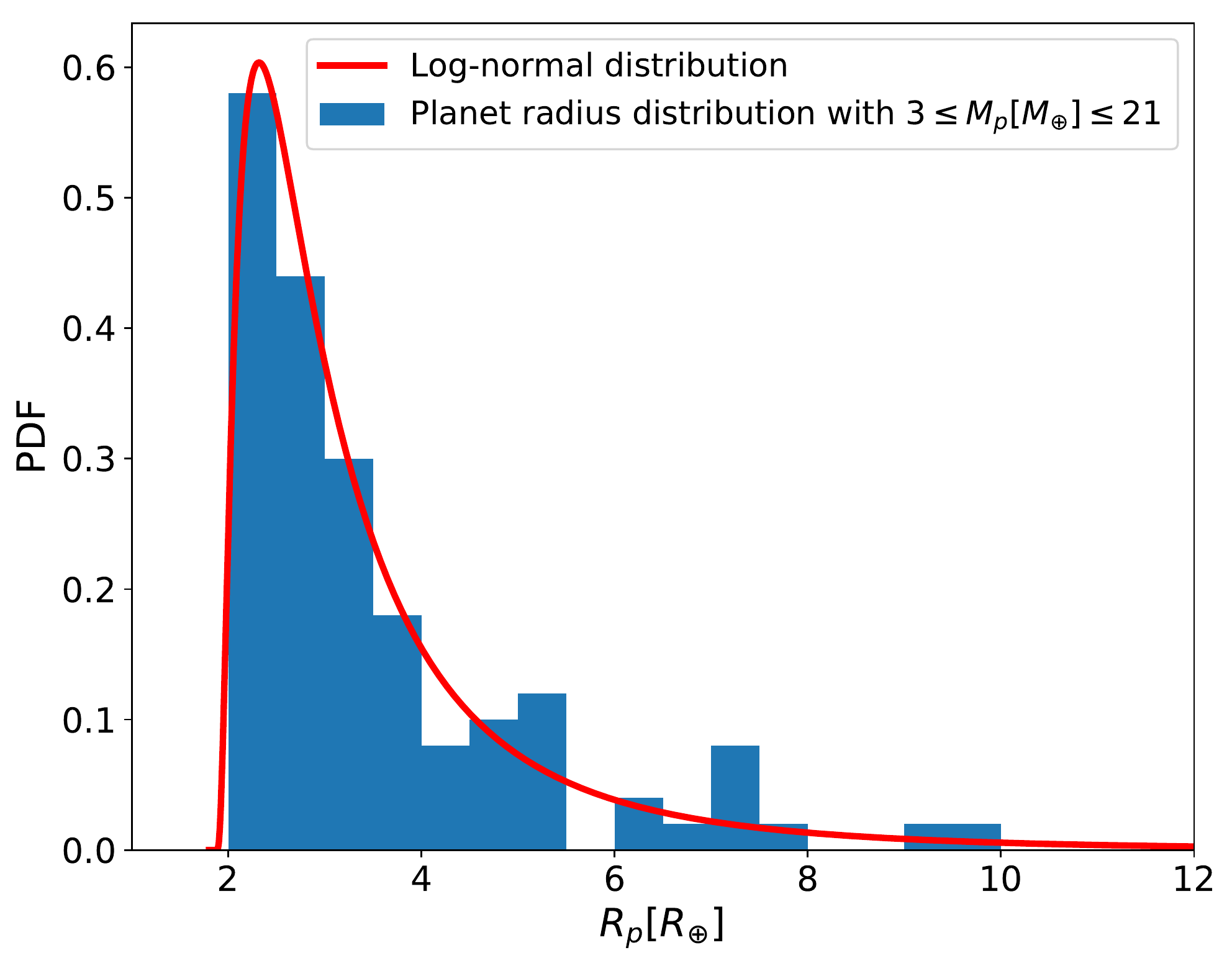}
\caption{Radius distribution of planets with masses within $3\sigma$ of the mass of HIP\,41378\,$f$ (obtained from NASA exoplanet archive) and fitted log-normal distribution used as prior on $R_{p}$.}
\label{rad_distr}
\end{figure}

\begin{figure*}[tph!]
 
    \includegraphics[width=0.67\linewidth]{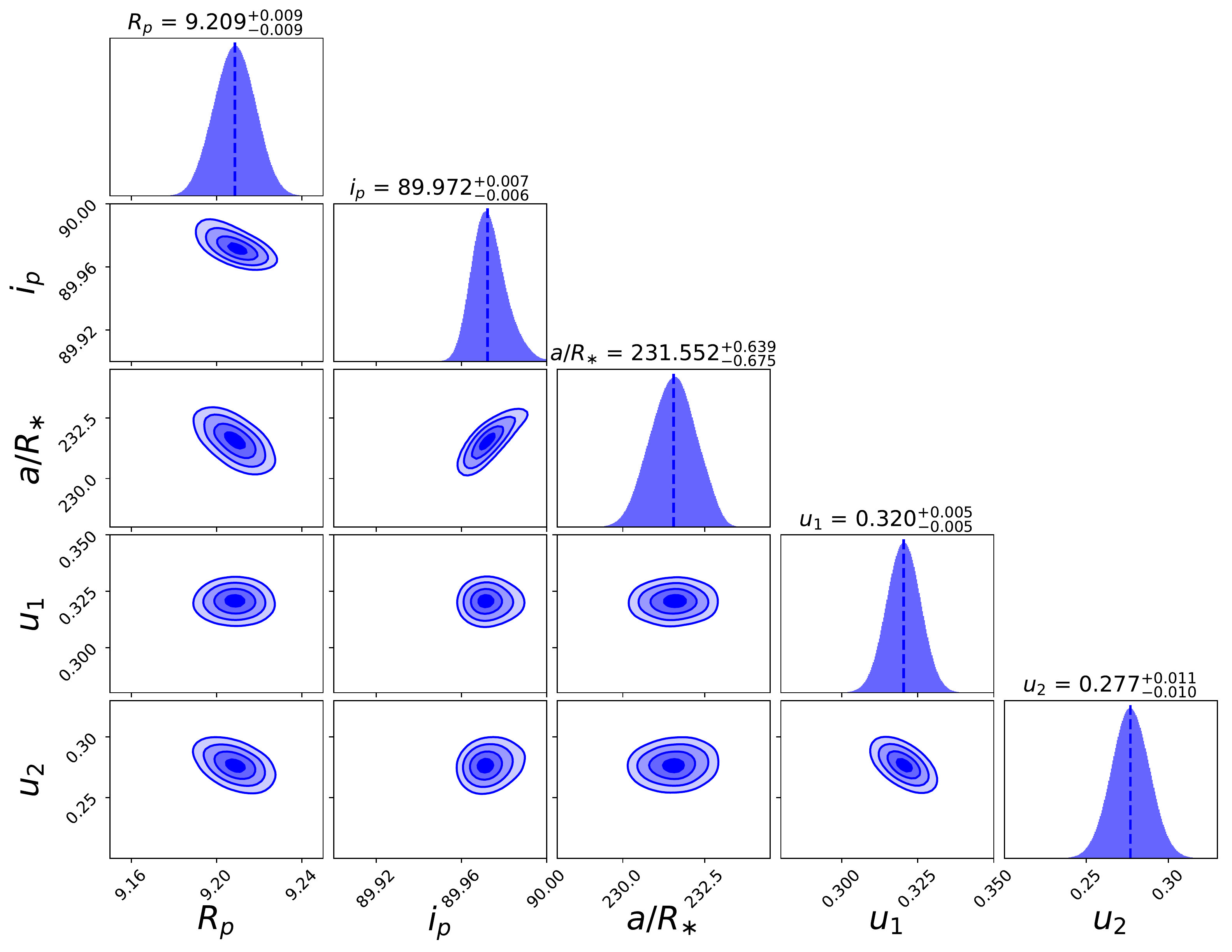}\\
    \includegraphics[width=1.02\linewidth]{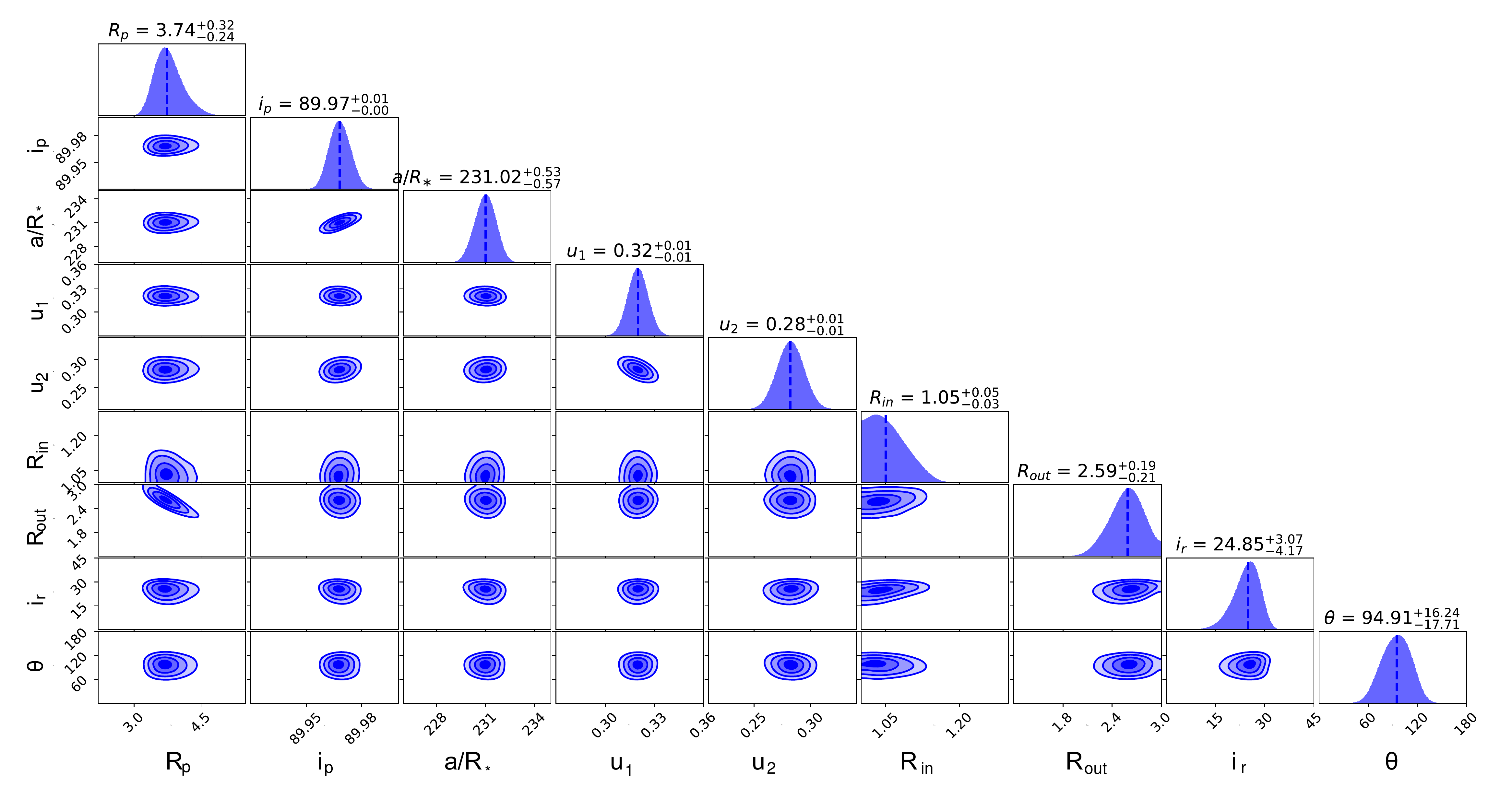}
    \caption{Posterior distribution of the planet-only model (top) and ringed planet model (bottom). The contours show the 0.5, 1, 1.5, and 2$\sigma$ uncertainties. The vertical lines show the medians of each parameter distributions and the quoted values are the medians and 68\% credible interval.}
    \label{posteriors}
\end{figure*}

\subsection{Computing evidence}
\label{evidence_calc}

Given some dataset $D$, a model $M$ with a set of parameters $\Theta$ has posterior probability determined from the Bayes rule as 

\begin{equation}
    P(\Theta|D,M)= \frac{P(D|\Theta,M)\,P(\Theta|M)}{P(D|M)} = \frac{\mathcal{L}(\Theta)\,\pi(\Theta)}{\mathcal{Z}}\, ,
\end{equation}
\noindent where $ \mathcal{L}(\Theta) = P(D|\Theta,M)$ is the likelihood, $\pi(\Theta) = P(\Theta|M)$ is the prior, and $\mathcal{Z}= P(D|M)$ is the evidence. We are interested in obtaining the evidence of each model in order to compare them. The evidence gives us a way to quantify the relative strength of each of the models given the data. It is computed as the integral over the entire prior domain, which makes it very sensitive to the choice of adopted priors. The evidence integral is given as
\begin{equation}
    \mathcal{Z} = \int \mathcal{L}(\Theta)\, \pi(\Theta)\, \mathrm{d}{\Theta}.
\end{equation}
We employ the \texttt{dynesty}\footnote{\href{https://dynesty.readthedocs.io/en/latest/index.html}{https://dynesty.readthedocs.io}} Python package \citep{speagle} which uses a nested sampling method \citep{skilling} to estimate the log evidence ($\log \mathcal{Z}$) by integrating the prior within nested contours of constant likelihood. A Gaussian likelihood function is used in our computation. The algorithm additionally provides posterior samples as a by-product. 

\subsection{Ring plane}
\label{ringplane}
The plane in which rings around a planet lie depends on the balance between the centrifugal force and stellar tide acting on the planet, which varies with the distance of the rings from the planet \citep{Tremaine_2009}. The distance from the planet where these forces balance out is defined as the Laplace radius $R_{L}$ given by \citep{sch11}
\begin{equation}
R_{L}^5=2J_2R_p^2a^3(1-e)^{3/2}\frac{M_p}{M_\ast}\,.
\end{equation}

\noindent Within $R_{L}$, rings settle in the equatorial plane of the planet, while beyond $R_L$ they settle in the orbital plane. Since rings spread out until $R_{Roche}$, it is straightforward to determine the ring plane by taking the ratio of $R_L$ and $R_{Roche}$ given by \citep{sch11}
\begin{equation}
\begin{split}
\label{laplace}
\frac{R_L}{R_{Roche}}\,\simeq\, 0.75\,\left(\,\frac{J_2}{0.01}\,\right)^{1/5}\,\left(\,\frac{M_p/M_{\ast}}{0.001}\,\right)^{-2/15}\,\left(\,\frac{R_p}{R_J}\,\right)^{2/5} \qquad \\
 \times \; \left(\,\frac{a/R_{\ast}}{21.5}\,\right)^{3/5}\left(\,\frac{\rho_r}{3g\,cm^{-3}}\,\right)^{1/3}\, ,
\end{split}
\end{equation}

\noindent where $J_{2}$ is the quadrupole moment of the planet (ranges from $\sim0.003$ for Uranus and Neptune to $\sim0.01$ for Jupiter and Saturn \citep{carterwinna}) and $R_{J}$ is the radius of Jupiter. For $R_L/R_{Roche} > 1$, the rings are entirely within $R_L$ and thus lie in the equatorial plane of the planet. For $R_L/R_{Roche} < 1$, rings extend beyond $R_L$ and thus transition from lying in the equatorial plane close to the planet to lying in the orbital plane  farther from the planet \citep{sch11}.

\subsection{Rossiter-McLaughlin}
\label{appendix_rm}

\begin{figure*}
\centering
\begin{minipage}[t]{.45\textwidth}
        \includegraphics[width=1\linewidth]{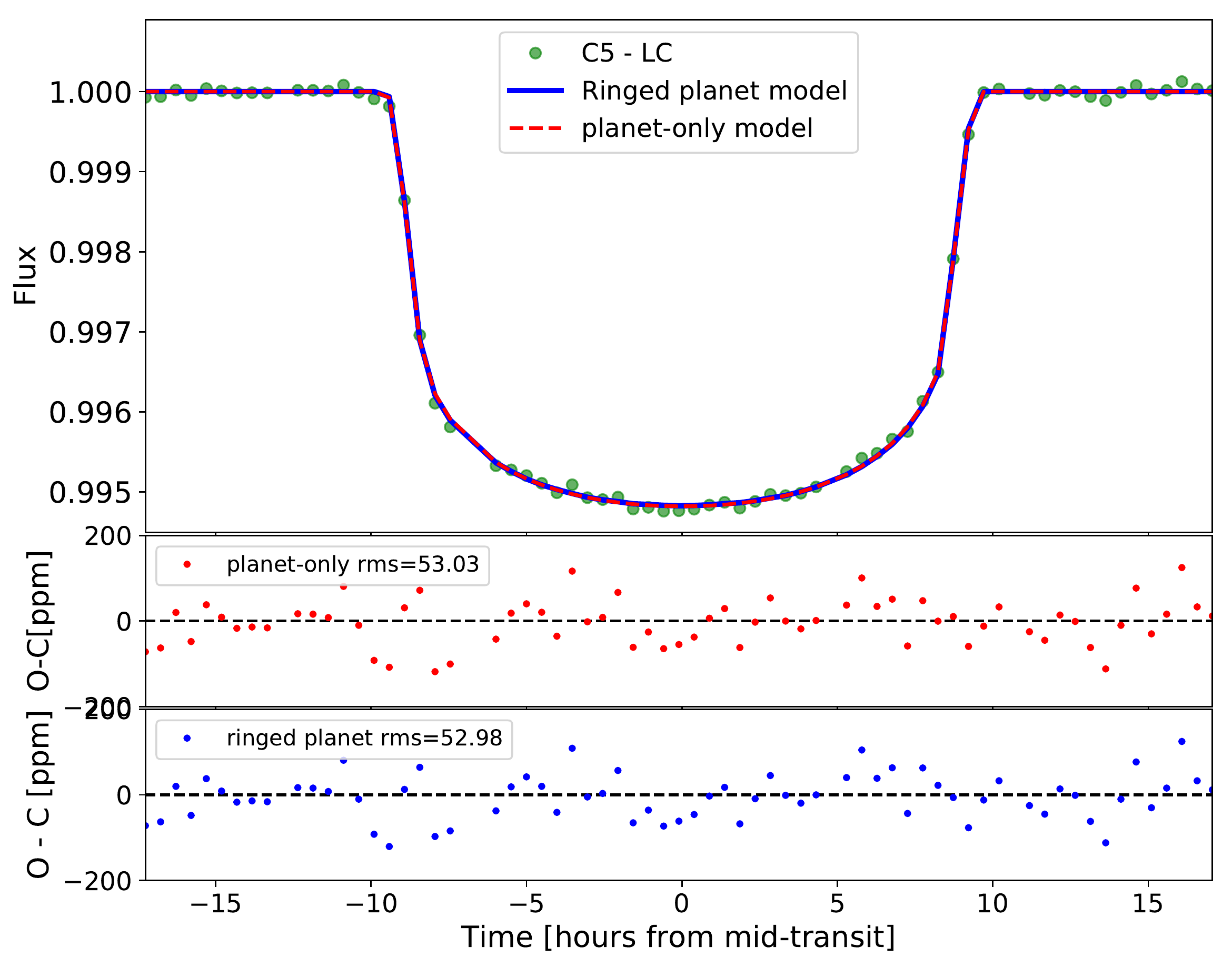}
        \caption{Fits of the planet-only model (red dashed line) and the ringed planet model (blue solid line) to the C5 LC data (green points). The residuals of same colour are also shown with the rms value in ppm.}
        \label{c5}\end{minipage}\qquad
\begin{minipage}[t]{.45\textwidth}
        \includegraphics[width=1\linewidth]{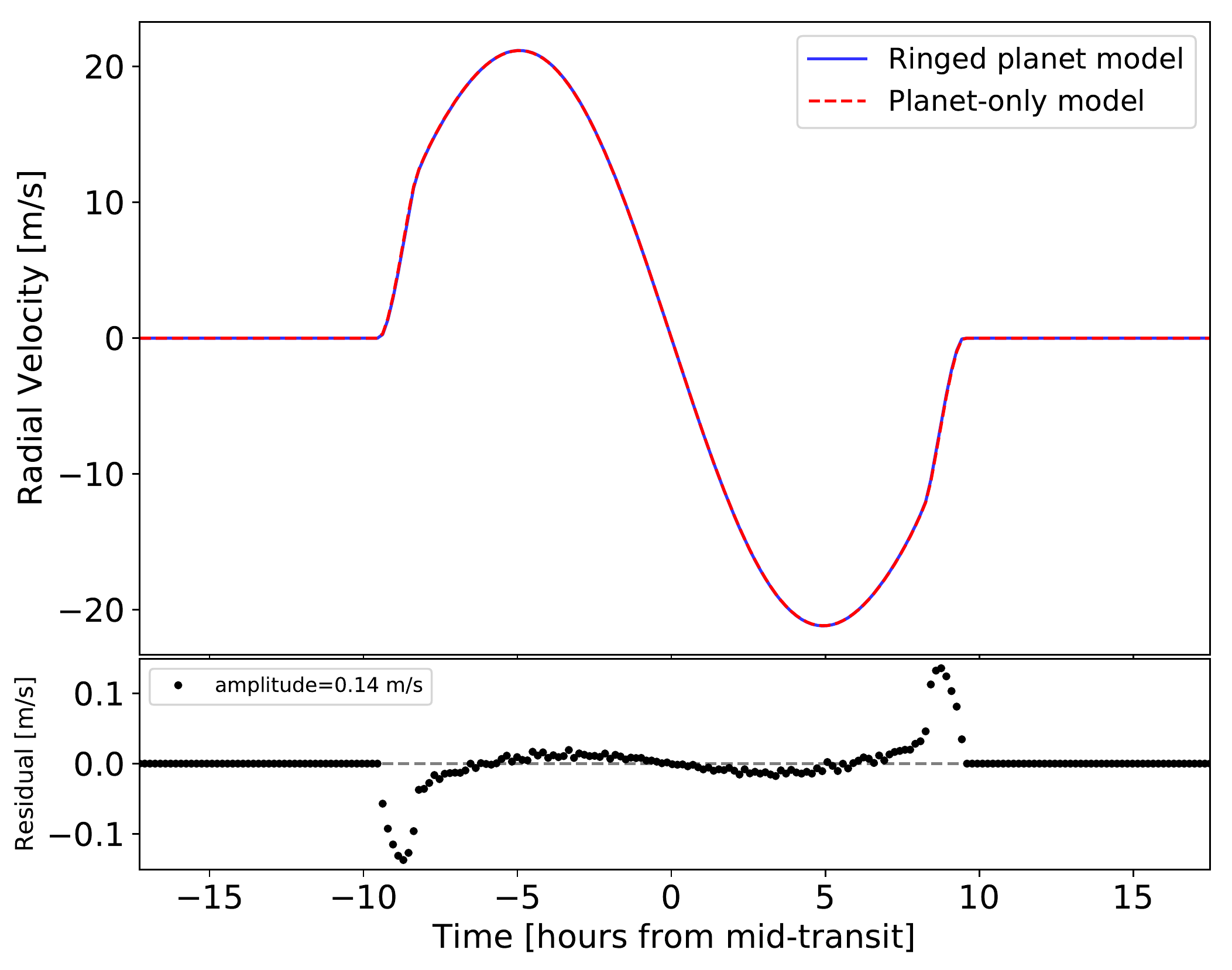}
        
        \caption{Comparison of expected RM signal of HIP\,41378\,$f$ using the ringed planet model (blue) and planet-only model (red dashed). The residual (bottom) between both models has an amplitude of 0.14\,m\,s$^{-1}$.}
        \label{rm}\end{minipage}
\end{figure*}

The RM measurements during transit can reveal the presence of rings around the planet. The rings affect the shape of absorption lines as the planet and ring cover different regions of the rotating star \citep{ohta,mooij}. The difference in the expected RM signal between the ringed planet model and the planet-only model given the projected stellar rotation velocity of 5.6\,km\,s$^{-1}$ is shown in Fig. \ref{rm}. The amplitude of the residual is only 0.14\,m\,s$^{-1}$, which might prove challenging even for the \texttt{ESPRESS0} \cite{pepe} spectrograph on the Very Large Telescope. However, with the long ingress duration, a long integration time can be used to attain high RV precision measurements of the ingress and egress. One of the RM methods \citep{mooij} involves resolving the distortions to the stellar line profile as the ring transits and does not require the entire transit to be observed. This makes it particularly useful for long period planet such as HIP\,41378\,$f$ with transits lasting longer than a night. 

\end{appendix}

\end{document}